\documentclass[3p]{elsarticle}
\pdfoutput=1

\usepackage[utf8]{inputenc}
\usepackage[T1]{fontenc}
\usepackage{uniinput}

\usepackage{hyperref}
\usepackage{amsmath,amssymb,amsfonts,bbm,dsfont,slashed,xfrac,nicefrac}

\DeclareMathOperator{\id}{id}
\newcommand{\1}{\ensuremath{\mathbbm{1}}}
\newcommand{\R}{\ensuremath{\mathds{R}}}
\newcommand{\tens}{\ensuremath{\otimes}}

\newcommand{\msbar}{\overline{\mbox{MS}}}
\DeclareMathOperator{\res}{res}
\newcommand{\momsnake}{\widetilde{\mbox{MOM}}}
\newcommand{\ren}{\raisebox{-0.2em}{$\raisebox{.2em}{$\phi$}_\textnormal{R}$}}
\newcommand{\frules}{\phi}

\newcommand{\FG}[3]{\raisebox{#2}{\includegraphics[height=#3]{#1}}}

\newcommand{\lessp}[2]{#1^{\includegraphics[trim=2 0 2 0, clip, height=2mm]{#2}}}
\newcommand{\ver}{\raisebox{-4.5pt}{\includegraphics[trim=2 0 2 0, clip, 
height=6mm]{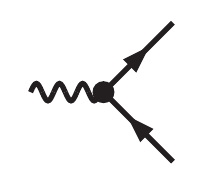}}}
\newcommand{\pho}{\includegraphics[trim=2 0 2 0, clip, 
height=3mm]{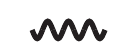}}
\newcommand{\phot}{\includegraphics[trim=2 0 2 0, clip, 
height=7mm]{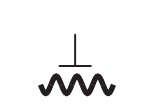}}
\newcommand{\phol}{\includegraphics[trim=2 0 2 0, clip, 
height=7mm]{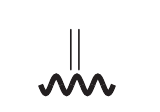}}
\newcommand{\pholzero}{\includegraphics[trim=2 0 2 0, clip, 
height=7mm]{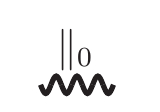}}
\newcommand{\pholone}{\includegraphics[trim=2 0 2 0, clip, 
height=7mm]{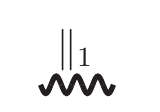}}
\newcommand{\fer}{\includegraphics[trim=2 0 2 0, clip, height=3mm]{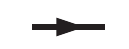}}

\begin{document}

\begin{frontmatter}

\title{Hopf-algebraic Renormalization of QED in the linear covariant Gauge}
\author{Henry Kißler}
\ead{kissler@physik.hu-berlin.de}
\address{Institut für Mathematik, Humboldt-Universität zu Berlin,\\ Rudower Chaussee 25, 
D-12489 Berlin, Germany}

\begin{abstract}
In the context of massless quantum electrodynamics (QED) with a linear covariant gauge fixing, the 
connection between the counterterm and the Hopf-algebraic approach to renormalization is examined.
The coproduct formula of Green's functions contains two invariant charges, which give rise to 
different renormalization group functions. All formulas are tested by explicit computations to 
third 
loop order. The possibility of a finite electron self-energy by fixing a generalized linear 
covariant gauge is discussed. An analysis of subdivergences leads to the conclusion that such a 
gauge only exists in quenched QED.
\end{abstract}


\end{frontmatter}

\section{Introduction}

Quantum electrodynamics is the first instance of a quantum gauge theory. The underlying gauge 
invariance serves both as guiding principle to construct more general theories as quantum 
chromodynamics or the standard model and as a symmetry, which restricts the physical predictions of 
the theory, e.g.\ the results in perturbation theory.

An omnipresent issue in perturbative calculations is the emergence of divergent integrals,
which require renormalization of the physical quantities. The Hopf-algebraic approach to 
renormalization provides a  mathematically rigorous, combinatorial treatment of divergences 
\cite{Kreimer:1997dp} and a graph-by-graph realization of the renormalization group equation 
\cite{Connes:1999yr,Connes:2000fe}. The description of quantum gauge theories has been extensively 
studied. Several articles explore the Hopf-algebraic structure in combination with combinatorial 
Dyson-Schwinger equations 
\cite{Kreimer:2005rw,Kreimer:2006ua,vanSuijlekom:2006ig,vanSuijlekom:2006fk,vanSuijlekom:2009zz} 
and its implication for physical quantities such as the beta function 
\cite{vanBaalen:2008tc,vanBaalen:2009hu,Klaczynski:2013fca}.

The perturbative treatment of a gauge theory requires a gauge fixing. The Landau gauge is most 
convenient for the usage of renormalization group techniques, because of the absence of 
renormalization effects. However, the generalized linear covariant gauge fixing introduces a new 
Lorentz structure, which requires a renormalization constant and contributes an additional 
expression to the renormalization group equation. Therefore, some aspects of the 
aforementioned articles implicitly require the Landau gauge. This article provides a diagrammatic 
description of the renormalization process of QED in the general linear covariant gauge and examines 
its implications on the structure of the Hopf algebra of Feynman graphs. Finally, these results and 
the freedom of choice of a gauge parameter are exploited to derive a restriction for the 
next-to-leading log terms in the self-energy of the electron.

The article is organized as follows. Section \ref{sec:renormalization} reviews the classical 
renormalization conditions with special emphasis on the gauge parameter. Its influence is exhibited 
in a diagrammatic description of the renormalization of the electron self-energy at second loop 
order in section \ref{sec:diaren}. These sections form the basis for the definition of the Hopf 
algebra structure and the description of Green's functions by combinatorial Dyson-Schwinger 
equations in section \ref{sec:hopfalgebra}. In section \ref{sec:higherorder}, gauge parameters of 
higher order in the coupling constant are introduced. This generalized linear covariant gauge is 
used to study the possibility of a finite electron self-energy.

\section{The class of linear covariant gauges}
\label{sec:renormalization}

Consider massless quantum electrodynamics (QED) in the linear covariant gauge; the Lagrangian is 
expressed in terms of renormalized fields and renormalization constants,
\begin{align}
  L_\text{QED} =  - \frac{1}{4} Z_3 F_{μν}F^{μν} + i Z_2 \bar{ψ} 
\slashed{\partial} ψ - e Z_1 
      \bar{ψ} \slashed{A} ψ -\frac{1}{2 ξ} Z_4 (\partial_μ A^μ)^2.
      \label{eq:qedLagrangian}
\end{align}
The gauge parameter $ξ$ parametrizes the covariant gauge fixing term, which breaks gauge 
invariance of the Lagrangian and contributes a linear term to the equations of motion.
The gauge fixing renormalization constant is termed $Z_4$, the other renormalization constants are 
labelled by the traditional convention. In the Lagrangian 
(\ref{eq:qedLagrangian}), the renormalization constants are absorbed by definition of the bare 
coupling, bare fields and bare gauge parameter:
\begin{gather}
 e_0 = \frac{Z_1}{Z_2 Z_3^{\nicefrac{1}{2}}} e, \quad ψ_0 = Z_2^{\nicefrac{1}{2}} ψ, \quad A^μ_0 = 
  Z_3^{\nicefrac{1}{2}} A^μ, \quad ξ_0 = \frac{Z_3}{Z_4} ξ.
  \label{eq:renormalization}  
\end{gather}
Throughout the paper, bare and renormalized quantities are denoted by the subscript zero and 
without a subscript, respectively. 
These renormalization constants are utilized to absorb divergent expressions. 
After quantization of the bare fields, the bare Green's functions, which depend on bare quantities, 
and the customary Feynman rules of quantum electrodynamics are deduced.
Three of the bare Green's functions require renormalization, namely 
the photon, the electron and the electron-photon-vertex Green's function. 

First, we concentrate on the photon Green's function and aspects pertaining to its renormalization. 
The introduction of the linear gauge fixing facilitates the calculation of the bare photon 
propagator, in the momentum space with external momentum $k$ it reads:
\begin{align}
 P^{μν} (k,ξ_0) = \frac{1}{k^2}\left( g^{μν} - \frac{k^μ k^ν}{k^2} \right) + ξ_0 \frac{k^μ 
  k^ν}{k^4}.
  \label{eq:photonProp}
\end{align}
The propagator decomposes into two Lorentz structures, a transversal part which vanishes 
after contraction with $k_μ$ or $k_ν$ and a longitudinal part which is proportional to $ξ_0$.
Gauge conditions can be imposed by fixing the value of the bare gauge parameter $ξ_0$, commonly 
used gauges are the transversal Landau gauge $ξ_0 = 0$ and the Feynman gauge $ξ_0 = 1$.
The bare Green's function of the photon $D_0^{μν}$ can be expressed as a geometric series in the 
bare photon self-energy $Π^{μν}_0$ which consists of all one-particle irreducible (i.e.\ 
2-edge-connected) Feynman graphs with two external photons. A direct analysis of quantum 
electrodynamics Feynman rules 
results in Ward's identities\footnote{These identities are sometimes called Ward-Takahashi 
identities. It should be noted that several people contributed to these identities and their 
generalizations, to name only a few: Ward \cite{Ward:1950xp}, Green \cite{Green:1953}, Takahashi 
\cite{Takahashi:1957xn}. Here, we refer to them as Ward's identities, as coined by 't Hooft in the 
general context of non-abelian gauge theories \cite{'tHooft:1971fh}.}
\begin{gather}
  k_μ Π^{μν}_0 (k) = 0 \quad \textnormal{and} \quad  k_ν Π^{μν}_0 (k) = 0.
\end{gather}
In other words, there is no longitudinal part in the photon self-energy and we can define the 
tensor reduced self-energy function
\begin{gather}
  Π^{μν}_0 (k) = T^{μν}(k) Π_0 (k^2), \quad T^{μν} (k) = g^{μν}k^2 - k^μ k^ν.
  \label{eq:wardidentities}
\end{gather}
The transversality of the self-energy simplifies the calculation of the bare Green's function, 
and as a result, the transversal part is multiplied by a factor of the geometric series of the 
self-energy function, but the longitudinal part of the propagator remains unmodified.
\begin{align}
  D^{μν}_0(k,ξ_0) = \frac{1}{k^2}\left(g^{μν}  - \frac{k^μ k^ν}{k^2} \right) \frac{1}{1 - Π_0(k^2)}
+ ξ_0 \frac{k^μ k^ν}{k^4}
\label{eq:photonGF}
\end{align}
This Green's function can be related to its renormalized equivalent by using equation 
(\ref{eq:renormalization}) and substituting the bare fields by renormalized fields in the 
time-ordered product of the field operators.
\begin{align}
 D^{μν} (k,α,ξ)  = \frac{1}{Z_3} D^{μν}_0 (k,α_0,ξ_0)
 \label{eq:bareAndRenPhoton}
\end{align}
This equation indicates that the renormalization constants introduce a scaling between bare and 
renormalized Green's functions, but do not influence the Lorentz structure, therefore also the 
renormalized Green's function satisfies the Ward's identities and obeys the same equation 
(\ref{eq:photonGF}) as the bare Green's function. A comparison of their Lorentz structures implies 
the following renormalization conditions:
\begin{align}
 & Π(k^2, α, ξ ) = Π_0 (k^2,α_0,ξ_0) + C_3 - C_3 Π_0 (k^2,α_0,ξ_0)  \\
 & Z_4 = 1
\end{align}
The first equation explains the renormalization of the photon Green's function by introduction of 
the counterterm $C_3$, for counterterms the convention $Z_i = 1 - C_i,\ i = 1,2,3,4$ is used. 
In combination with equation (\ref{eq:renormalization}), the latter condition characterizes the 
renormalization of the gauge parameter. Consequently, the gauge parameter renormalization 
constant is solely determined by the renormalization of the photon self-energy, i.e. by the 
renormalization constant of the photon self-energy, $Z_3$.
\begin{align}
  ξ_0 = Z_3 ξ
\end{align}
Due to the appearance of the renormalization scale in $Z_3$, the renormalized gauge parameter 
becomes dependent on the renormalization scale $μ$. Effectively, this has no influence on the 
renormalization of the photon self-energy and the beta function; these observables are independent 
of the gauge parameter. However, the renormalization scale dependence of the gauge parameter results 
in an additional term in the renormalization group equations.
\begin{align}
 \left( -\partial_L + β\, α \partial_α + δ\, ξ \partial_ξ + γ^r\right) G^r(L,α,ξ) = 0, \quad
r\in\left\{\fer{},\pho{},\ver{}\right\}
\end{align}
The renormalization group functions $β$ and $δ$ respectively describe the renormalization 
scale dependence of the coupling and the gauge parameter, $L = \log (\nicefrac{-k^2}{μ^2})$ denotes 
a kinematic variable and $γ^r$ is called the anomalous 
dimension of the Green's function of type $r$. These functions can be defined 
by differentiating with respect to the scale $μ^2$.
\begin{gather}
  \partial_L = - μ^2 \partial_{μ^2}, \quad α β(α,ξ) = μ^2 \frac{d α}{d μ^2}, \quad γ^r (α,ξ) = 
- μ^2 \frac{d\ln Z_r}{dμ^2}, \quad α δ(α,ξ) =  μ^2\frac{d ξ}{d μ^2}
\end{gather}
We will show how the renormalization group functions and equations arise from a subdivergence 
analysis of Green's functions using Hopf-algebraic methods in paragraph \ref{sec:sstary}. The 
renormalization of the gauge parameter has of cause no effect on the QED beta function, which is 
determined by the photon self-energy and hence gauge parameter independent. However, as the electron 
self-energy has a non-trivial gauge dependence, we study its influence on the renormalization 
process in the subsequent paragraphs.

\section{Diagrammatic renormalization of the electron self-energy}
\label{sec:diaren}

In the following, we work out the renormalized self-energy of the electron at two loops, which is 
the lowest order involving effects from the renormalization of the gauge parameter. A 
comparison between Zimmermann's forest formula and the renormalization conditions of 
one-particle irreducible Green's functions determines the counterterm $C_2$ at two loops.
This provides a graphical description which indicates how a single Feynman graph contributes to the 
counterterms---this is the crucial step for understanding the relation between counterterms and 
BPHZ as well as Hopf-algebraic renormalization.

\subsection{BPHZ renormalization via Zimmermann's forest formula}

A renormalization prescription turns mathematically ill-defined divergent Feynman integrals into 
well-defined expressions. Here, we utilize Zimmermann's forest formula \cite{Zimmermann:1969jj} 
i.e.\ the BPHZ renormalization prescription. The forest formula allows to lift the renormalization 
procedure of Feynman integrals to the level of Feynman graphs. In this sense, the whole 
renormalization process is understood as a combinatorial manipulation of Feynman graphs. For every 
individual Feynman graph $Γ$, this prescription guarantees a finite renormalized expression 
$\ren(Γ)$, which is defined by means of Zimmermann's forest formula
\begin{align}
  \ren (Γ) = \sum_{f\in \mathcal{F}_Γ} (-1)^{|f|}\left(\prod_{γ\in f}T(γ)\right) \frules(Γ/f).
 \label{eq:forestformula} 
\end{align}
The sum runs over all forest $f$ (a set of divergent one-particle irreducible subgraphs of $Γ$ 
which are either nested or disjoint). $T(γ)$ denotes the divergent part of the Feynman 
integral corresponding to $γ$. The divergent part of a Feynman integral can be defined by extending 
the integral (regularization) or by evaluating its external kinematic variables at a special point. 
In the first case, the operator $T$ might also extract some arbitrary finite part, fixing this 
freedom corresponds to the choice of a renormalization scheme. The Feynman integral corresponding to 
$\frules(Γ/f)$ is obtained by contracting all components of the forest $f$ in $Γ$. It is worth 
noting that contraction is sensitive to the Lorentz structure of a subgraph. The following example 
illustrates this issue.
\begin{align}
 \FG{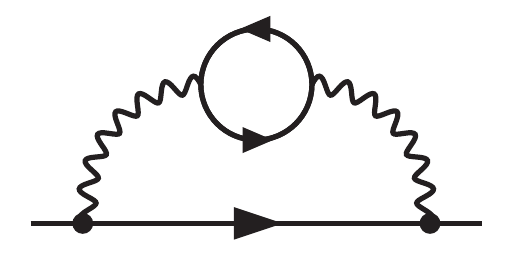}{-.5em}{2em} / \FG{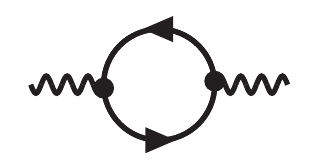}{-.75em}{2em} = \FG{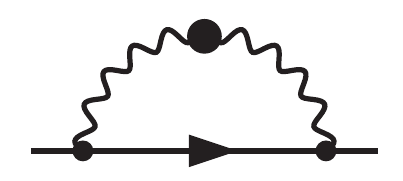}{-.5em}{2em} = 
\FG{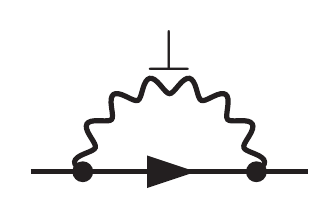}{-.55em}{2.5em}
 \label{eq:contractionEx}
\end{align}
This example demonstrates the contraction of the first order photon self-energy graph, which is 
inserted into the first order electron self-energy graph. In the first equation, the photon 
subgraph is contracted to a point, which represents the remaining Lorentz structure of the photon 
self-energy: the transversal tensor $T_{μν}(k) = g_{μν}k^2 - k^μ k^ν$. This Lorentz structure 
projects both the left and the right photon propagators on their transversal parts and cancels one 
of them in the second equation. Finally, the transversal part (i.e.\ Landau gauge term) of the 
first order electron self-energy remains. This result might also be anticipated from a calculation 
of the second order graph. An explicit evaluation of its photon subdivergence yields a purely 
transversal result. Hence, in an evaluation of the second order graph, only the transversal parts of 
the photon propagators contribute, therefore the contraction should also yield a purely transversal 
result.

\subsection{Renormalization conditions}

As demonstrated in the preceding paragraph, a transition from bare to renormalized 
one-particle irreducible Green's functions implies the following renormalization conditions.
\begin{align}
\label{eq:renormalizationCondition1}
 &\textnormal{vertex} &&Λ^ν(q_1,q_2,α,ξ,μ) = Λ_0^ν(q_1,q_2,α_0,ξ_0) -C_1 Λ_0^ν(q_1,q_2,α_0,ξ_0)\\
\label{eq:renormalizationCondition2}
 &\textnormal{electron} &&Σ(q,α,ξ,μ) = Σ_0(q,α_0,ξ_0) +C_2 -C_2 Σ_0(q,α_0,ξ_0)\\
 &\textnormal{photon} &&Π(q,α,ξ,μ) = Π_0(q,α_0,ξ_0) +C_3 -C_3 Π_0(q,α_0,ξ_0)
 \label{eq:renormalizationCondition3}
\end{align}
Primary, these conditions expose that renormalization is carried out in two steps: first a 
substitution of bare parameters (which carry the  subscript $0$) to renormalized parameters is 
performed and second the remaining divergences are absorbed by defining appropriate counterterms. 
Bare and renormalized parameters are connected by a product of renormalization constants, which also 
depend on the coupling or gauge parameter. Therefore, the substitution replaces bare parameters by 
series of renormalized parameters. In case of the bare coupling parameter and its product with the 
gauge parameter, the expansion in the renormalized coupling parameter up to second order reads
\begin{align}
α_0 & = \frac{Z_1^2}{Z_2^2 Z_3} α = α \left(1 + (-2C_1+2C_2+C_3)|_1 + O(α^2)\right)\\
ξ_0 α_0 & = \frac{Z_1^2}{Z_2^2Z_3} Z_3 ξ α = ξ α \left( 1+ (-2C_1+2C_2)|_1 + O(α^2)\right).
\end{align}
At first order, bare parameters are directly rewritten in terms of renormalized parameters, e.g.\
$α_0 = α + O(α^2)$ for the coupling parameter. This observation simplifies the derivation of 
one-loop counterterms. However, at two loops, the coupling renormalization requires all one-loop 
counterterms.

\subsection{One-loop counterterms and renormalized parameters}

The one-loop counterterms are determined by the one-loop one-particle irreducible graphs, which 
have no subdivergences, therefore only two forests occur in Zimmermann's forest formula: the empty 
graph and the full graph, which determines the counterterm. A comparison with the renormalization 
conditions (\ref{eq:renormalizationCondition1})-(\ref{eq:renormalizationCondition3}) fixes the 
signs.
\begin{align}
  &Λ_0 = \FG{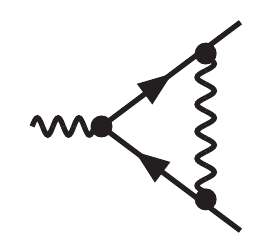}{-1em}{2.5em} &&C_1 = T\left[\FG{v1}{-1em}{2.5em}\right]\\
  &Σ_0 = \FG{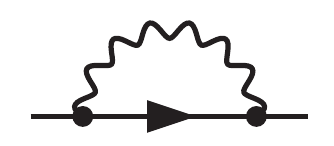}{-.5em}{1.5em} &&C_2 = -T\left[\FG{e1}{-.5em}{1.5em}\right]\\
  &Π_0 = \FG{p1}{-.75em}{2em} &&C_3 = -T\left[\FG{p1}{-.75em}{2em}\right]
\end{align}
As described for the forest formula (\ref{eq:forestformula}), the operator $T$ extracts the 
divergent part of a Feynman graph and represents a particular renormalization scheme. Now, the 
bare coupling parameter is rewritten as a series in the renormalized coupling parameter.
\begin{align}
 α_0 &= \frac{Z_1^2}{Z_2^2Z_3} α= α \left(1 + T\left[-2\FG{v1}{-1em}{2.5em} -2\FG{e1}{-.5em}{1.5em} 
-\FG{p1}{-.75em}{2em}\right] +O(α^2) \right)
\intertext{Notice that the bare coupling parameter occurring in the Feynman graphs is tacitly 
substituted by the renormalized coupling, this introduces higher order terms which do not 
contribute 
at second order. Beside the coupling constant, a Feynman graph might also include factors of the 
gauge parameter $ξ_0$. The product of both contributes the following renormalization terms.}
  ξ_0 α_0 &= \frac{Z_1^2}{Z_2^2Z_3} Z_3 ξ α = ξα \left(1 + T\left[-2\FG{v1}{-1em}{2.5em} 
-2\FG{e1}{-.5em}{1.5em} 
\right] +O(α^2) \right)
\end{align}
It should be remarked that Ward's identity $Z_1 = Z_2$ simplifies the renormalization of the gauge 
and coupling parameters---the product $ξα = ξ_0 α_0$ is not renormalized. However, these 
simplifications might be misleading in the comparison with subdivergences of the forest formula 
and are therefore not taken into account here.

The different substitution rules for the parameters  $α$ and $ξα$ make it necessary to 
distinguish between longitudinal and transversal photons. Therefore, the photon propagator is 
divided into the  sum of its transversal and longitudinal parts, which are denoted by the labels 
$\perp$ and $\parallel$, respectively.
\begin{align}
\label{eq:photonsplit}
\begin{tabular}{ccccc}
 $P^{μν} (k, ξ_0)$ & $=$ & 
 {$\!{\begin{aligned}\frac{1}{k^2}\left( g^{μν} - \frac{k^μ k^ν}{k^2} \right)\end{aligned}}$} & 
$+$ & 
{$\!{\begin{aligned}ξ_0 \frac{k^μ k^ν}{k^4}\end{aligned}}$}\\
$\FG{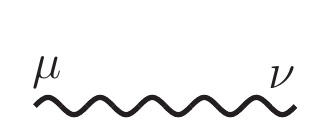}{-.2em}{2em}$ & $=$ & $\FG{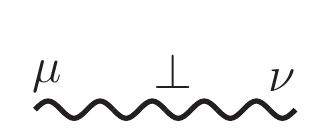}{-.2em}{2em}$ & $+$ & $\FG{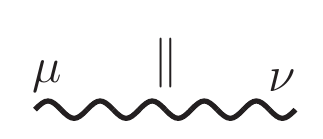}{-.2em}{2em}$
\end{tabular}
\end{align}
In this notation, Feynman graphs are build from purely transversal and purely longitudinal photon 
propagators---all possible combinations need to be taken into account; e.g.\ the bare self-energy 
at first loop order becomes a sum of two graphs.
\begin{align}
 Σ_0|_1 = \FG{e1}{-.5em}{1.5em} = \FG{e1t}{-.5em}{2em} + \FG{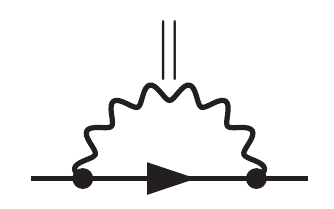}{-.5em}{2em}
\end{align}

\subsection{Two-loop renormalization of the self-energy of the electron}

In this paragraph, the renormalized electron self-energy and its counterterm $C_2$ are worked out  
at second loop order.  From the renormalization condition (\ref{eq:renormalizationCondition2}) and 
the previous discussion five contributions are expected.
\begin{align}
 Σ(α,ξ)|_2 = Σ_0|_2 + Σ_0|_1(α,ξ) + C_2|_1(α,ξ) -C_2|_1 Σ_0|_1 + C_2|_2
\end{align}
The first term contains all two-loop graphs from the bare self-energy, the second and third term 
arise from the substitution of bare by renormalized parameters in the first order bare self-energy 
and its counterterm, the fourth term is determined by the one-loop bare self-energy and its 
counterterm, the fifth term is the two-loop counterterm, which is constructed in the following.

The bare self-energy of the electron at two loops reads
\begin{align}
 Σ_0|_2 = \FG{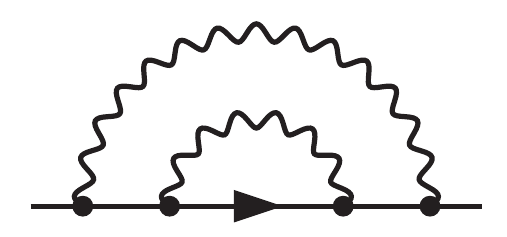}{-.5em}{2em} + \FG{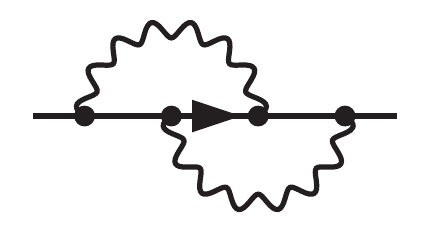}{-1.25em}{2em} + \FG{e23}{-.5em}{2em}.
\end{align}
In these two-loop graphs, all parameters are tacit substituted by renormalized parameters, this 
causes contributions at three and higher loops, but no contribution at second loop order. However, 
the parameter substitution in the one-loop graphs contribute at second order.
\begin{align}
\label{eq:asymmetry}
  Σ_0|_1 (α,ξ) = - T \left[2 
\FG{v1}{-1em}{2.5em}+2\FG{e1}{-.5em}{1.5em}+\FG{p1}{-.75em}{2em}\right]\FG{e1t}{-.5em}{2em}
- T \left[2\FG{v1}{-1em}{2.5em}+2\FG{e1}{-.5em}{1.5em}\right]\FG{e1l}{-.5em}{2em}
\end{align}
Here, the renormalization of the gauge parameter protects the longitudinal part of the self-energy 
from divergences of the transversal photon self-energy, this is in accordance with the contraction 
of subdivergences in the forest formula, as exemplified in  (\ref{eq:contractionEx}). The same 
applies to the first order counterterm.
\begin{align}
  C_2|_1 (α,ξ) = -T\left[Σ_0|_1 (α,ξ) \right]
\end{align}
The fourth contribution is a product of the one-loop self-energy and its counterterm, again 
tacitly rewritten in terms of renormalized parameters.
\begin{align}
   -C_2|_1 Σ_0|_1 = T\left[\FG{e1}{-.5em}{1.5em}\right]\FG{e1}{-.5em}{1.5em}
\end{align}
Finally, a comparison between the derived terms and forests from Zimmermann's formula determines 
the counterterm of the electron self-energy at two loops.
\begin{align}
  C_2|_2 = - T\left[\FG{e21}{-.5em}{2em} + \FG{e22}{-1.25em}{2em} + \FG{e23}{-.5em}{2em}\right]
  - T\left[T\left[\FG{e1}{-.5em}{1.5em}\right]\FG{e1}{-.5em}{1.5em}\right]
\end{align}
Notice that a composed term of the first order graph contributes. This term arises due to 
the fact that the counterterm is defined as the difference of terms generated by the 
forest formula \eqref{eq:forestformula} and terms resulting from the renormalization of parameters 
(as the coupling or gauge parameter). For the sake of completeness, we also provide the 
renormalization constant at two loops.
\begin{align}
 Z_2 = 1 + T\left[\FG{e1}{-.5em}{1.5em}\right]
 +T\left[\FG{e21}{-.5em}{2em}+\FG{e22}{-1.25em}{2em}+\FG{e23}{-.5em}{2em}\right]
 +T\left[ T\left[ \FG{e1}{-.5em}{1.5em}\right]\FG{e1}{-.5em}{1.5em}\right] + O(α_0^3)
\end{align}
As a non-trivial check, we substituted the graphs by their dimensional regularized results in 
$4-2ε$ dimensions and reproduced the well-known result of $Z_2$ \cite{Grozin:2007zz}.

\section{Hopf-algebraic renormalization of QED}
\label{sec:hopfalgebra}

The purpose of this paragraph is to examine the effect of the linear covariant gauge fixing on
the Hopf algebra of quantum electrodynamics and to perform the program of Hopf-algebraic 
renormalization. This involves the definition of the Hopf algebra of QED 
Feynman graphs, the construction of Green's functions by combinatorial Dyson-Schwinger equations 
(DSE), the derivation of a coproduct formula for the Green's functions, and the 
evaluation of their renormalized expressions. Moreover, the derivation of the renormalization 
group equation by means of the Dynkin operator $S\star Y$ is discussed.

\subsection{Hopf algebra structure of massless QED in the linear covariant gauge}

The construction of the Hopf algebra of QED is well-known \cite{Kreimer:1997dp,vanSuijlekom:2006ig}.
Nonetheless, we recall its definitions and basic properties to establish our conventions. In the 
previous paragraph, we have observed that renormalization distinguishes transversal and 
longitudinal 
parts of the photon self-energy. Therefore, both Lorentz structures are distinguished by assigning 
the different graph labellings $\perp$ and $\parallel$ to the photon propagators. The labelled 
propagators represent the edge types of a Feynman graph.

Let $H$ be the free commutative algebra over $\R$ generated by the set of all divergent 
one-particle irreducible Feynman graphs together with the product $m:H\tens H \rightarrow H$. The 
unit $\1\in H$ is identified with the empty graph and the homomorphism $u:\R \rightarrow H$ with 
$u(1) = \1$ denotes the unit map. A Feynman graph is build from propagators and vertices of the set
\begin{align}
\label{eq:propsandvertices}
    \mathcal{R}_\textnormal{QED} \in \left\{ \ver{}, \fer{}, \phot{}, \phol{}
\right\}.
\end{align}
Further, a Feynman graph is divergent if its external leg structure matches an element of 
$\mathcal{R}_\textnormal{QED}$. Note that according to equation \eqref{eq:photonsplit}, the photon 
propagator is rewritten in terms of the transversal and the longitudinal propagator. A generic 
photon self-energy graph contributes to both of these Lorentz structures, the projection onto one 
of these Lorentz structures is denoted by assigning either the $\perp$ or the $\parallel$ label to 
the external legs of the Feynman graph. In case of a vertex and an electron self-energy graph, the 
tacit projection onto their divergent Lorentz structures ($γ_μ$ and $\slashed{q}$) is always 
understood.

In \cite{Kreimer:1997dp} Kreimer showed that a coproduct $Δ : H \rightarrow H \tens H$ can be 
defined by
\begin{align}
  Δ Γ = \sum_{γ \trianglelefteq Γ} γ \tens Γ/γ, \quad Γ\in H.
\end{align}
Where the summation goes over all products of disjoint one particle irreducible divergent 
subgraphs of $Γ$ (including both the empty and the full subgraph) and $Γ/γ$ denotes the 
Feynman graph obtained by replacing all components of $γ$ by their external structures.

The counit $ε:H \rightarrow \R$ is the homomorphism which satisfies $ε(\1) = 1$ and vanishes on the 
complement of $\R\1$. These definitions yield a bialgebra $(H,m,Δ,u,ε)$, which possesses a 
grading induced by the loop number of a Feynman graph. Hence, this bialgebra is indeed a Hopf 
algebra \cite{sweedler1969hopf}, its antipode $S:H\rightarrow H$ is recursively defined by
\begin{align}
 S(Γ) = - Γ - \sum_{\substack{γ \triangleleft Γ\\ γ \neq \1}} S(γ) Γ/γ, \quad Γ \in H.
\end{align}
As pointed out in \cite{Brown:2011pj}, this recursion is solved by a sum over all forests which 
exclude the full graph $Γ$.
\begin{align}
 S(Γ) = - Γ - \sum_f (-1)^{|f|}γ_f\tens Γ/γ_f, \quad γ_f = \prod_{γ \in f} γ 
\end{align}
This formula reveals a striking similarity to Zimmermann's forest formula \eqref{eq:forestformula}. 
Indeed, Zimmermann's forest formula is reproduced by the convolution of the Feynman rules $\frules$ 
and a twisted version of the antipode.
\begin{align}
\label{eq:renFR}
 \ren (Γ) = m\circ (\frules \circ S_T \tens \frules )\circ Δ(Γ)
\intertext{Here, $S_T$ is the antipode twisted with the renormalization scheme operator $T$}
 S_T (Γ) = - T\circ \left( Γ + \sum_{\substack{γ \triangleleft Γ\\ γ \neq \1}}S_T(γ) Γ/γ\right), 
\quad Γ \in H.
\end{align}
In this formulation, the finiteness of the renormalized expression $\ren$  is understood in terms 
of an algebraic Birkhoff decomposition \cite{Manchon:2001bf,Kreimer:2012nk}. It is worth 
emphasising that the proof of this theorem clarifies why the renormalization scheme operator $T$ 
has 
to satisfy the Rota Baxter equation
\begin{align}
\label{eq:rotabaxter}
T(γ_1) T(γ_2) = - T(γ_1 γ_2) + T\circ \left( T(γ_1)γ_2 + γ_1 T(γ_2) \right),
\end{align}
a necessary condition of a well-defined renormalization scheme that was previously anticipated by 
practitioners \cite{Collins:1984xc}.

Equation \eqref{eq:renFR} provides a prescription to renormalize an individual Feynman graph. 
However, the renormalization conditions 
\eqref{eq:renormalizationCondition1} to \eqref{eq:renormalizationCondition3} and the 
corresponding counterterms refer to one particle irreducible Green's functions. Therefore, the next 
topic in our discussion is the relation between the coproduct and one particle irreducible Green's 
functions.

\subsection{Green's functions from combinatorial DSE}

The coproduct extracts by definition subdivergences of individual Feynman graphs, its application 
on one-particle irreducible Green's function is best understood in the language of combinatorial 
DSE. 

In \cite{Broadhurst:2000dq} and \cite{Kreimer:2005rw} Broadhurst and Kreimer demonstrated how 
one-particle irreducible Green's functions are constructed as solutions of combinatorial DSE. In 
this language, a Green's function is built by insertion of subdivergences into skeleton graphs 
(Feynman graphs which are free of subdivergences). More precisely, for a given skeleton graph $γ$ 
they defined an insertion operator $B_+^γ$, which takes a product of Feynman graphs as argument and 
maps it to the sum of all possible insertions of these Feynman graphs into the skeleton $γ$ 
multiplied by some combinatorial factor, defined in \cite{Kreimer:2005rw}. 

\begin{figure}[t]
\caption{\label{fig:skeletons}Low order examples of skeleton graphs. As indicated by the first two 
graphs, each labelling of photon propagators induces a skeleton graph, these labels are understood 
in the subsequent graphs.}
\begin{center}
 \FG{e1t}{-.5em}{2em}
 \FG{e1l}{-.5em}{2em}
 \FG{p1}{-.75em}{2em}
 \FG{v1}{-1em}{2.5em}
 \FG{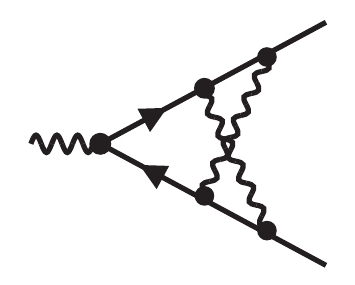}{-1.2em}{3em}
 \end{center}
\end{figure}

To describe QED in the linear covariant gauge, the vertex Green's function is denoted by 
$\lessp{X}{verline}$ and the electron self-energy by $\lessp{X}{ferline}$. The photon self-energy 
is 
represented by the Green's functions $X^\perp$ and $X^\parallel$ which respectively include the 
contributions to the transversal and the longitudinal Lorentz structure. With this definition, 
$X^\perp$ is only inserted into the transversal photon propagators and $X^\parallel$ only into the 
longitudinal photon propagators. The QED system of Dyson-Schwinger equations reads as follows.
\begin{align}
\label{eq:dseVerElePer}
 X^r & = \1 \pm \sum_{\substack{γ \textnormal{ skeleton}\\ \res(γ) = r}} B^γ_+ 
 \left(X^r Q_\perp^{n_\perp (γ)+1}Q_\parallel^{n_\parallel (γ)}\right)
 \quad \textnormal{for } r \in \left\{ \ver, \fer, \perp \right\}\\  
\label{eq:dseparallel}
 X^\parallel & = \1 - \sum_{\substack{γ \textnormal{ skeleton}\\ \res(γ) = \parallel}}
 B^γ_+ \left(X^\parallel Q_\perp^{n_\perp (γ)}Q_\parallel^{n_\parallel (γ)+1}\right)
\end{align}
Each of these sums go over all one particle irreducible skeleton graphs of the external leg 
structure $\res (γ)$; the vertex function contains an infinite number of skeleton graphs---some 
examples are provided in figure \ref{fig:skeletons}.
The number of transversal and longitudinal photon propagators of a Feynman graph $γ$ is denoted by 
$n_\perp(γ)$ and $n_\parallel(γ)$. In \eqref{eq:dseVerElePer}, all propagators receive the negative 
sign, the positive sign only applies for the series vertex graphs. 
The input of the insertion operators is written in terms of the invariant charges
\begin{align}
\label{eq:invariantcharges}
 Q_\perp = \frac{(\lessp{X}{verline})^2}{(\lessp{X}{ferline})^2X^\perp} \quad \textnormal{and} \quad
 Q_\parallel = \frac{(\lessp{X}{verline})^2}{(\lessp{X}{ferline})^2 X^\parallel}.
\end{align}
Notice that the argument of each insertion operator is defined such that every vertex of the 
skeleton is dressed by a factor $\lessp{X}{verline}$, every electron propagator by a factor 
$\nicefrac{1}{\lessp{X}{ferline}}$, every transversal and every longitudinal photon propagator by
a factor $\nicefrac{1}{X^\perp}$ and $\nicefrac{1}{X^\parallel}$ , respectively. The inverse 
of a one particle irreducible propagator series corresponds to the series of connected propagator 
graphs; a one particle irreducible Green's function is build from one particle irreducible vertex 
insertions, but requires insertions of connected propagator graphs.

The advantage of the combinatorial Dyson-Schwinger equations approach is that a properly defined 
insertion operator $B_+$ yields a well-behaved compatibility relation with the coproduct 
\cite{Kreimer:2005rw,Yeats:2008zy},
\begin{align}
\label{eq:hochschildcocycle}
  Δ \circ B_+ = B_+ \tens \1 + (\id \tens B_+ )\circ Δ.
\end{align}
This relation allows inductive proofs by induction on the number of subdivergences of a Feynman 
graph and implies a coproduct formula for Green's functions.

\subsection{The coproduct of QED Green's functions}

A closed formula for the coproduct on Green's functions has first been provided by Yeats 
\cite{Kreimer:2006ua}. By usage of the compatibility relation \eqref{eq:hochschildcocycle} Yeats 
derived a formula for the coproduct of one particle irreducible Green's functions in the case of a 
single invariant charge. However, it should be noted that all her proofs canonically generalize to 
systems of DSE involving multiple invariant charges by promoting the exponent of the single 
invariant charge to a multi-index; e.g. in case of the electron self-energy $n \equiv (n_\perp, 
n_\parallel)$ and
\begin{align}
 Q^n \equiv Q_\perp^{n_\perp} Q_\parallel^{n_\parallel}.
\end{align}
In addition to that, direct proofs of a coproduct formula for one particle irreducible Green's 
functions has been provided in \cite{vanSuijlekom:2006fk,Borinsky:2014xwa}. All these results imply 
the following coproduct formula for an one particle irreducible Green's function $X^r$.
\begin{align}
\label{eq:qedcoproduct1}
 & Δ X^r =  \sum_{0\leq n_\parallel \leq n} X^r Q_\perp^{n - n_\parallel}Q_\parallel^{n_\parallel} 
  \tens X^r_{n;n_\parallel} 
  \quad \textnormal{for } r \in \left\{ \ver, \fer, \perp \right\}\\  
\label{eq:qedcoproduct2} 
 & Δ X^\parallel =  \sum_{0\leq n_\perp \leq n} X^\parallel Q_\perp^{n_\perp} 
Q_\parallel^{n-n_\perp} \tens X^\parallel_{n;n_\perp} 
\end{align}
In this formula, $X^r_{n;n_\perp}$ denotes all Feynman graphs of the Green's function $X^r$ 
which have $n$ loops and $n_\perp$ transversal photon propagators; and $X^r_{n;n_\parallel}$ 
analogously with restriction to $n_\parallel$ photon propagators.
In this formulation, the exponent shift of $n_\perp$ in \eqref{eq:dseVerElePer} and of $n_\parallel$ 
in \eqref{eq:dseparallel} was absorbed into the loop number $n$. Also note that the sum of both 
exponents of the invariant charges equals the loop number of a cograph, which appear on the 
right side of the tensor product.

\subsection{Renormalization group equation by analysis of subdivergences}
\label{sec:sstary}

If one restricts oneself to the one-scale case, i.e.\ zero momentum transfer at the interaction 
vertex, all renormalized Green's functions can be expanded in a single variable $L$ which denotes 
the kinematic log terms.
\begin{align}
 G^r(L,α,ξ) = \ren \left(X^r\right) = 1+\sum_{n\geq 1} γ^r_n(α,ξ) L^n
\end{align}
The coefficients of this expansion are called anomalous dimensions, whereas $γ_1^r$ is the 
actual physical anomalous dimension. For the general case of an arbitrary number of scales the 
reader is referred to \cite{Brown:2011pj}. 

In \cite{Panzer:2012gp} 
Panzer showed that for the case of one scale Feynman graphs 
and a proper renormalization scheme (e.g.\ the $\momsnake$ scheme), the Birkhoff decomposition 
implies that the renormalized Feynman rules map $\ren$ is a morphism of bialgebras and hence 
Feynman rules, or more precisely the anomalous dimensions, can be rewritten in terms of a 
combinatorial operation on the Hopf algebra of Feynman graphs.
\begin{align}
\label{eq:sstary}
 γ_n^r = \frac{1}{n!} σ^{\star n}(X^r) \quad \textnormal{where} \quad σ = \ren \circ Y^{-1} \circ 
(S\star Y)
\end{align}
The product $S\star Y$ is defined as the convolution $m \circ (S \tens Y) \circ Δ$ and $σ^{\star n}$ 
denotes the convolution of $n$ maps $σ$. For further results and properties of $σ$ and the Dynkin 
operator $S\star Y$, the reader is referred to \cite{Yeats:2008zy,Panzer:2014kia} and references 
therein. Here, we only mention that the linear map $S\star Y$ is a projection and vanishes 
on products of Feynman graphs. These  properties greatly simplify the evaluation of $σ$ on products 
of combinatorial Green's functions.
\begin{align}
\label{eq:anomalousdim}
 & σ (X^r) = γ^r \\
\label{eq:betafct}
 & σ (Q_\perp) = \left( 2σ(\lessp{X}{verline}) - 2σ(\lessp{X}{ferline}) -  σ(X^\perp) \right) = β\\
\label{eq:deltafct}
 & σ (Q_\parallel) = \left( 2σ(\lessp{X}{verline}) - 2σ(\lessp{X}{ferline}) - σ(X^\parallel) \right) 
  = β + δ
\end{align}
In the first equation, the application on a combinatorial Green's function $X^r$, $σ$ extracts its 
anomalous dimension $γ^r$, in the second equation the invariant charge $Q^\perp$ yields a sum of 
invariant charges, its result is denoted by the function $β(α,ξ)$, and the third equation defines an 
additional function $δ(α,ξ)$. Note that these functions represent the QED renormalization group 
functions, which simplify on account of the Ward-Takahashi identity $\lessp{γ}{verline} = 
\lessp{γ}{ferline}$ and the transversality of the photon $γ^\parallel = 0$ as follows.
\begin{align}
 β & = 2\lessp{γ}{verline} - 2\lessp{γ}{ferline} - γ^\perp = - γ^\perp\\
 δ & = γ^\perp - γ^\parallel = γ^\perp
\end{align}
In fact, these functions appear in the renormalization group equation, which is now derived in the 
case of the electron self-energy. First, the $L^n$ coefficient is expressed in terms of the 
projection $σ$. Then, the coproduct formula \eqref{eq:qedcoproduct1} is applied.
\begin{align}
(n+1) \lessp{γ}{ferline}_{n+1} &= \frac{σ^{\star (n+1)}}{n!} (\lessp{X}{ferline}) = σ \star 
\frac{σ^{\star n}}{n!} (\lessp{X}{ferline})\\
  & = \sum_{0\leq n_\parallel \leq n} σ(\lessp{X}{ferline} Q_\perp^{n-n_\parallel} 
  Q_\parallel^{n_\parallel}) \frac{σ^{\star n}}{n!} 
\left(\lessp{X}{ferline}_{n;n_\parallel}\right)\\
\intertext{Due to the fact that $σ$ is a linear map and vanishes on products of Feynman graphs, 
products of Green's functions turn into sums and the exponents of the invariant charges become 
factors.}
 (n+1) \lessp{γ}{ferline}_{n+1} & = \sum_{0\leq n_\parallel \leq n} (γ_r + n β + n_\parallel δ) 
\frac{σ^{\star n}}{n!}	\left(\lessp{X}{ferline}_{n;n_\parallel}\right)
\end{align}
In the next step, the renormalization group functions are identified through 
(\ref{eq:anomalousdim}-\ref{eq:deltafct}). Note that the grading coefficients $n$ and $n_\parallel$ 
can be replaced by $α \partial_α$ and $ξ \partial_ξ$ acting on $σ^{\star n} 
(\lessp{X}{ferline}_{n;n_\parallel})$. As result, the Green's function coefficient 
$\lessp{γ}{ferline}_{n+1}$ is related to coefficient $\lessp{γ}{ferline}_n$, the anomalous 
dimension $\lessp{γ}{ferline}$ and the renormalization group functions $β$ and $δ$. This 
restricts the renormalized Green's function to satisfy the renormalization group equation.
\begin{align}
  0 = (-\partial_L + β α \partial_α + δ ξ \partial_ξ + \lessp{γ}{ferline}) \lessp{G}{ferline}
\end{align}
It is worth emphasising that the appearance of the two renormalization group functions $β$ and $δ$ 
is directly linked to the presence of the two invariant charges in the coproduct formula of one 
particle irreducible Green's function (\ref{eq:qedcoproduct1}-\ref{eq:qedcoproduct2}).

\subsection{Hopf algebra for the practitioner}

In this paragraph, all previously derived result concerning the Hopf-algebraic renormalization 
of QED in the linear covariant gauge are tested by explicit computations. We use {\sc Qgraf} to 
generate all one particle irreducible Feynman graphs to three loops and list them in table 
\ref{tab:graphs}. 
\begin{table}[t]
\label{tab:graphs}
\begin{tabular}{cccc}
\hline
Green's function & one loop & two loops & three loops \\
\hline
$\ver$ & 1 & 9 & 100\\
$\fer$ & 1 & 3 & 20\\
$\pho$ & 1 & 3 & 20\\
\hline
\end{tabular}
\caption{Number of Feynman graphs for each one-particle irreducible Green's function.}
\end{table}
The vertex Green's function is considered at zero photon momentum transfer. This 
reduces the problem to an evaluation of two point graphs, which was performed with {\sc Form} 
\cite{Vermaseren:2000nd} in combination with the {\sc Mincer} package \cite{Larin:1991fz}. The 
transversality of the photon propagator and the Ward-Takahashi identity served as a check of 
consistency of the applied methods. The renormalization has been carried out utilizing the 
Hopf-algebraic renormalization \eqref{eq:renFR} and the coproduct formula \eqref{eq:qedcoproduct1} 
within both the $\msbar$ and $\momsnake$ scheme. In the former scheme, we reproduced self-energy of 
the photon \cite{Gorishnii:1991hw}.
This provides a proof of concept of the above Hopf-algebraic formulas 
(\ref{eq:qedcoproduct1}-\ref{eq:qedcoproduct2}) and \eqref{eq:renFR}. Here, we state our result of 
the self-energy of the electron in the $\momsnake$ scheme as this is of interest in the next 
paragraph.
\begin{multline}
\label{eq:renselfenergy}
 Σ(α,ξ) = ξL\left(\frac{α}{4π}\right)
 + \left(-\frac{1}{2}ξ^2L^2 - (\frac{3}{2}+2n_f)L\right)\left(\frac{α}{4π}\right)^2\\
 + \left(\frac{1}{6}ξ^3L^3+\left((\frac{3}{2}+2n_f)ξ-2n_f-\frac{8}{3}n_f^2\right)L^2+(\frac{3}{2}
-2n_f+\frac { 8 } { 3}n_f^2)L\right)\left(\frac{α}{4π}\right)^3
\end{multline}

\section{Higher order gauge parameters}
\label{sec:higherorder}

In \cite{Johnson:1959zz}, Johnson and Zumino asserted that all divergences in the self-energy of 
the electron can be eliminated by a suitable choice of gauge, resulting in a finite wave-function 
renormalization constant $Z_2$ and, by Ward's identity, a finite vertex renormalization constant 
$Z_1$.\footnote{The crucial ingredient is the exact knowledge of the behaviour of the electron 
self-energy under gauge transformations. This has been studied by Landau and Khalatnikov 
\cite{Landau:1955zz} and Zumino \cite{Zumino:1959wt}.} For quenched quantum electrodynamics (QED 
without insertions of photon self-energy graphs), such a gauge has been explicitly constructed by 
Baker, Johnson, and Willey in \cite{Johnson:1964da}, where finite solutions of the self-energy were 
derived by solving a system of truncated Dyson-Schwinger equations. To achieve this, they had to 
allow for a coupling dependent gauge parameter:
\begin{align}
\label{eq:bjwgauge}
  ξ_0(α_0) = \frac{3}{2}\frac{α_0}{4π}
\end{align}
Their technique of introducing gauge parameters of higher orders in the coupling parameter was also
applied in quenched quantum electrodynamics in the first evaluation of the three-loop beta function 
by Rosner \cite{Rosner:1966zz} and in Broadhurst's calculation of the anomalous dimensions of the 
quenched theory to four loops \cite{Broadhurst:1999he}. These results motivate us to examine the 
behaviour of these coupling dependent gauge parameters under renormalization and which kind of 
divergences can be cancelled by a suitable choice of gauge. We promote the bare gauge parameter to 
a series in the bare coupling parameter.
\begin{align}
  ξ_0(α_0) = \sum_{n\geq 0} ξ_0^{(n)}α_0^n
\end{align}
A gauge fixing corresponds to a specific choice of the parameters $ξ_0^{(n)}$, which are required 
to be free parameters. However, as we have seen in the previous paragraphs, after renormalization, 
the gauge parameter becomes a power series in the coupling parameter and its coefficients are 
determined through the renormalization condition $ξ_0 = Z_3\, ξ$. To avoid this kind of 
contradiction, we also introduce renormalized higher order gauge parameters $ξ_0^{(n)} = Z_3^{n+1} 
ξ^{(n)}$ and replace the renormalized gauge parameter by a series of renormalized gauge parameters.
\begin{align}
ξ (α) = \sum_{n\geq 0} ξ^{(n)}α^n
\end{align}
This definition maintains the renormalization condition $ξ_0 (α_0) = Z_3 ξ(α)$, which is necessary 
due to the fact that the self-energy of the photon is transversal. The photon propagator is 
modified by a series of longitudinal parts of higher order in the gauge parameter.
\begin{align}
\label{eq:photonprophigherorder}
  P^{μν} (k,ξ_0) = \frac{1}{k^2}\left( g^{μν} - \frac{k^μ k^ν}{k^2} \right) + ξ_0^{(0)} \frac{k^μ 
  k^ν}{k^4} + ξ_0^{(1)}α_0 \frac{k^μ k^ν}{k^4} + \cdots
\end{align}
In this generalized linear covariant gauge, Feynman graphs are build from vertices and edges of 
the infinite set
\begin{align}
 \label{eq:propsandvertices2}
    \mathcal{R} \in \left\{ \ver, \fer, \phot, \pholzero, \pholone, \cdots \right\},
\end{align}
where the transversal part of the photon propagator is denoted by the $\perp$ label and the 
longitudinal term which is proportional to $ξ_0^j$ is denoted by the label $\parallel_j$.

First, the quenched sector of QED in this gauge is discussed. The vertex and electron one particle 
irreducible Green's functions are solutions of the following system of DSE.
\begin{align}
 Δ X^r = \1 + \sum_{\substack{γ \textnormal{ skeleton}\\ \res(γ) = r}} B_+^γ (X^r Q^{|γ|}), 
\quad r \in  \left\{ \ver, \fer \right\}
\end{align}
Where the invariant charge of the quenched theory is defined as
\begin{align}
 Q = \frac{(\lessp{X}{verline})^2}{(\lessp{X}{ferline})^2}.
\end{align}
Note that these DSE only differ from \eqref{eq:dseVerElePer} of the full theory 
by the property that no photon self-energy graphs are inserted into the skeletons. Again, the 
application of \eqref{eq:hochschildcocycle} implies a closed formula for the coproduct of the one 
particle irreducible Green's function in the quenched sector.
\begin{align}
  \lessp{X}{ferline} = \sum_{n\geq 0} \lessp{X}{ferline} Q^n \tens \lessp{X}{ferline}_n
\end{align}
Here, $\lessp{X}{ferline}_n$ denotes all $n$-loop Feynman graphs of the quenched Green's function 
$\lessp{X}{ferline}$. It should be remarked that the same formula follows from the coproduct 
formula of the full theory by dividing out the Hopf-ideals generated by 
$X^\perp_{n_\perp,n_\parallel}$ and $X^\parallel_{n_\perp,n_\parallel}$.
As demonstrated in the previous paragraph, this coproduct formula in combination with 
\eqref{eq:sstary} restricts the $L$ expansion of the electron self-energy. The absence of 
transversal photon subdivergences implies that the quenched invariant charge vanishes under $σ = 
\ren \circ Y^{-1} \circ \left( S\star Y \right)$.
\begin{align}
 σ (Q) =  2\lessp{γ}{verline} - 2\lessp{γ}{ferline} = 0
\end{align}
This corresponds to the fact that coupling parameter $α$ and the gauge parameter $ξ$ are not 
renormalized in the vertex and electron Green's function of the quenched theory and yields the 
following simplified renormalization group equation.
\begin{align}
  (\partial_L - \lessp{γ}{ferline}) (1-Σ_\textnormal{quenched}) = 0, \quad \textnormal{with }
  \lessp{γ}{ferline} = - \partial_L Σ_\textnormal{quenched}(0)
\end{align}
This ordinary differential equation determines the self-energy of the electron by means of its 
anomalous dimension, which is up to a sign the linear log term of the self-energy. Hence, the 
unique solution reads
\begin{align}
 Σ_\textnormal{quenched} = 1 - \exp \left(L \lessp{γ}{ferline} \right).
\end{align}
In this way, the question of a vanishing quenched self-energy of the electron is converted to a 
vanishing anomalous dimension. Indeed, there is a choice of higher order gauge parameters such that 
the anomalous dimension vanishes. Recall that the one-loop self-energy graph is only sensitive 
to the longitudinal part of the photon propagator (see \eqref{eq:renselfenergy}). Therefore, the 
self-energy of the electron vanishes in the Landau gauge $ξ^{(0)} = 0$ at first loop order. Now, 
observe that the one-loop graph contributes at higher loop orders by the coupling dependent 
parts of the photon propagator \eqref{eq:photonprophigherorder}. Moreover, this shift of the first 
order graph yields a linear log term which is proportional to a unspecified gauge parameter. In 
other words, the linear log term of the electron self-energy at loop order $n$ can be cancelled by 
fixing the value of the gauge parameter $ξ^{(n-1)}$. In this gauge fixing, the anomalous dimension 
vanishes.
\begin{align}
 Σ_\textnormal{quenched}(α, \tilde{ξ}(α)) = 0 \quad \textnormal{ for some} \quad \tilde{ξ}(α) = 
\sum_{j\geq 0} \tilde{ξ}^{(j)} α^j
\end{align}
A vanishing renormalized self-energy implies a finite bare self-energy. In other words, the 
constructed gauge fixing cancels all divergences and allows a finite renormalization constant 
$Z_2$, which is the original statement of Baker, Johnson, and Willey.

Finally, the unquenched case of QED is discussed. As demonstrated in the analysis of the quenched 
sector, the cancellation of the linear log terms of the self-energy already determines the full set 
of higher order gauge parameters $\tilde{ξ}(α)$. This in combination with the three-loop result 
\eqref{eq:renselfenergy} determines the gauge to second order.
\begin{align}
\label{eq:gaugeparameterho}
 \tilde{ξ}(α) =  0 + \left( \frac{3}{2}+2n_f\right) α + \left(-\frac{3}{2}+2n_f-\frac{8}{3}n_f^2 
  \right) α^2
\end{align}
Note that the quenched limit $n_f = 0$ coincides with the results of Baker, Johnson, and Willey 
\eqref{eq:bjwgauge} at two loops and Broadhurst at three loops \cite{Broadhurst:1999he}. However, in 
the full theory, the cancellation of the anomalous dimension of the 
electron does not imply a vanishing self-energy and at third order a quadratic log term remains:
\begin{align}
 Σ(α, \tilde{ξ}(α) ) = \left( -2n_f-\frac{8}{3}n_f^2\right)L^2 \left(\frac{α}{4π}\right)^3.
\end{align}
A $L$ dependent term arises from a divergent subgraph in Zimmermann's forest formula. Therefore, 
this nonvanishing $L$ term corresponds to a remaining subdivergence in the self-energy of the 
electron. This subdivergence is necessarily cancelled by a divergent counterterm. Hence, the higher 
order gauge parameters of the generalized linear covariant gauge fixing are not sufficient to remove 
all divergences from the self-energy of the electron, or equivalently to provide a finite 
renormalization constant $Z_2$.

It is interesting to note that the pure existence of this gauge technique restricts the 
next-to-leading log term in the self-energy of the electron within the original linear covariant 
gauge. More precisely, a next-to-leading log term proportional to 
\begin{align}
\label{eq:forbidden}
 ξ^0 n_f^0 L^{n-1} \left( \frac{α}{4π}\right)^n, \quad n\geq 3
\end{align}
is forbidden. 

In other words, a Feynman graph which contributes to the next-to-leading log term 
possesses 
\begin{itemize}
 \item (at least) a longitudinal photon propagator
 \item or (at least) a closed fermion loop. 
\end{itemize}
This follows from the fact that such a term contradicts the existence of higher order gauge 
parameters which induce the vanishing of the quenched self-energy: Notice that a term of form 
\eqref{eq:forbidden} is a quenched contribution and recall that the leading-log term allows a 
direct evaluation via truncated DSE and reads
\begin{align}
\label{eq:nloopllog}
  (-1)^{n+1} \frac{ξ^nL^n}{n!} \left( \frac{α}{4π}\right)^n, \quad n\geq 1.
\end{align}
Now, replace the gauge parameter $ξ$ by the coupling dependent series $ξ(α)$, whose coefficients 
are fixed by the requirement that the electron anomalous dimension vanishes. As argued before, this 
guaranties a vanishing electron self-energy. Due to an expansion of the gauge parameter series, a 
log term of a particular order contributes also at higher loop orders. As a result, the 
next-to-leading log term at $n+1$ loops receives a contribution from the leading-log term at $n$ 
loops \eqref{eq:nloopllog}. Further, all terms which depend on the gauge parameter vanish because of 
the vanishing coefficient $ξ^{(0)} = 0$ of \eqref{eq:gaugeparameterho}. The only remaining 
term is of the form \eqref{eq:forbidden}. However, by construction of the series $ξ(α)$, all log 
terms (including the next-to-leading log term) of the quenched sector vanish. Hence, the 
next-to-leading log term of the electron self-energy vanishes in the Landau gauge ($ξ$ has 
non-zero exponent) or is not within the quenched sector ($n_f$ has non-zero exponent).

\section{Conclusion}

In the analysis of the renormalization conditions of QED in the linear covariant gauge, we 
stressed the fact that the gauge parameter $ξ$ is renormalized, which distinguishes the transversal 
and the longitudinal part of the photon propagator in the process of renormalization. This 
dissimilarity was examined in the renormalization of the self-energy of the electron at second 
order (see e.g.\ \eqref{eq:asymmetry}) and also appeared in the Hopf algebra structure---the 
coproduct of one particle irreducible Green's functions 
incorporates two invariant charges \eqref{eq:invariantcharges}. Moreover, we demonstrated that each 
invariant charge translates into a renormalization group function and reproduces the well-known 
renormalization group equations \cite{Pascual:1984zb}. Our analysis supports the interpretation that 
every renormalized parameter is represented by an invariant charge in the coproduct formula. It 
might be the topic of future work to extend this analysis to a non-linear gauge, where Ward's 
identities are expected to relate particular invariant charges, as observed in the context of 
non-abelian gauge theories \cite{vanSuijlekom:2006fk}.

Algebraic Birkhoff decomposition was utilized to construct renormalized Feynman rules by means of 
the antipode $S$ and without requiring any reference to renormalization constants and counterterms, 
this revealed the similarity to Zimmermann's forest formula i.e.\ the BPHZ renormalization 
prescription. For the example of the renormalization constant $Z_2$, we demonstrated how a 
graph-by-graph representation of the renormalization constants is constructed by comparing the 
renormalization conditions with the structure of subdivergences which are characterized by the 
forest formula or the coproduct formula. As non-trivial check of all these Hopf-algebraic results, 
QED was renormalized in the linear covariant gauge to third loop order.

Finally, the extension of the linear covariant gauge by gauge parameters of higher order in the 
coupling parameter was analysed. Following the approach of Baker, Johnson, and Willey 
\cite{Johnson:1964da}, we constructed a gauge fixing for quenched QED which implied a vanishing
self-energy and hence a finite renormalization constant $Z_2$. However, it was demonstrated 
that such a gauge fixing does not exist in the full theory. It should be remarked that our 
conclusion coincides with a renormalization group analysis of Adler and Bardeen 
\cite{Adler:1971pm}, their argument is restricted to quenched QED and fails if photon propagator 
insertions are not neglected. Nonetheless, the pure existence of such a gauge technique implied 
that the next-to-leading log term in Landau-gauged quenched QED vanishes.

\section*{Acknowledgments}

The author is indebted to D.\ Kreimer for his patience, constant encouragement to finish
this article, and for bringing the topic of higher order gauge parameters to his attention. He also
thanks J.M.\ Bell and J.A.\ Gracey for helpful discussions and advice concerning the usage 
of {\sc Form}, as well as D.J.\ Broadhurst for stimulating discussions and pointing out the 
paper \cite{Broadhurst:1999he}. Further thanks to L.\ Klaczynski for proofreading the manuscript. 
The figures in this paper were created with {\sc jaxodraw} \cite{Binosi200476} and the {\sc 
axodraw} package \cite{Vermaseren:1994je}. A DAAD scholarship is gratefully acknowledged.

\bibliographystyle{elsarticle-num}
\bibliography{linQED}

\end{document}